\documentclass[
aps,
twocolumn, % twocolumn onecolumn
showpacs,
prl, % prl preprint reprint,
superscriptaddress] %,floatfix]
{revtex4-1}
\usepackage{graphicx}
\usepackage{color}
\usepackage{ulem}
\usepackage{verbatim}
\usepackage{amsmath}
\usepackage{amssymb}
\usepackage{hyperref}
\usepackage{epstopdf}
\usepackage{float}
\hypersetup{
  colorlinks   = true, %Colours links instead of ugly boxes
  urlcolor     = blue, %Colour for external hyperlinks
  linkcolor    = blue, %Colour of internal links
  citecolor    = blue %Colour of citations
}

\begin{document}
\title{A cold atom temporally multiplexed quantum memory with cavity-enhanced noise suppression}%

\author{Lukas Heller}
\email[Contact: ]{lukas.heller@icfo.eu}
\affiliation{ICFO-Institut de Ciencies Fotoniques, The Barcelona Institute of Science and Technology, 08860 Castelldefels (Barcelona), Spain}

\author{Pau Farrera}
\email[Contact: ]{pau.farrera@icfo.eu}
\altaffiliation{Present address: Max-Planck-Institut für Quantenoptik,  Hans-Kopfermann-Strasse  1,  85748  Garching,  Germany}
\affiliation{ICFO-Institut de Ciencies Fotoniques, The Barcelona Institute of Science and Technology, 08860 Castelldefels (Barcelona), Spain}

\author{Georg Heinze}
\altaffiliation{Present address: TOPTICA Projects GmbH, Lochhamer Schlag 19, 82166 Gräfelfing, Germany}
\affiliation{ICFO-Institut de Ciencies Fotoniques, The Barcelona Institute of Science and Technology, 08860 Castelldefels (Barcelona), Spain}

\author{Hugues de Riedmatten}
\homepage{http://qpsa.icfo.es}
\affiliation{ICFO-Institut de Ciencies Fotoniques, The Barcelona Institute of Science and Technology, 08860 Castelldefels (Barcelona), Spain}
\affiliation{ICREA-Instituci\'{o} Catalana de Recerca i Estudis Avan\c cats, 08015 Barcelona, Spain}%

\begin{abstract}

Future quantum repeater architectures, capable of efficiently distributing information encoded in quantum states of light over large distances, will benefit from multiplexed photonic quantum memories. In this work we demonstrate a temporally multiplexed quantum repeater node in a laser-cooled cloud of $^{87}$Rb atoms. We employ the DLCZ protocol where pairs of photons and single collective spin excitations (so called spin waves) are created in several temporal modes using a train of write pulses. To make the spin waves created in different temporal modes distinguishable and enable selective readout, we control the dephasing and rephasing of the spin waves by a magnetic field gradient, which induces a controlled reversible inhomogeneous broadening of the involved atomic hyperfine levels. We demonstrate that by embedding the atomic ensemble inside a low finesse optical cavity, the additional noise generated in multi-mode operation is strongly suppressed. By employing feed forward readout, we demonstrate distinguishable retrieval of up to 10 temporal modes. For each mode, we prove non-classical correlations between the first and second photon. Furthermore, an enhancement in rates of correlated photon pairs is observed as we increase the number of temporal modes stored in the memory. The reported capability is a  key element of a quantum repeater architecture based on multiplexed quantum memories.
\end{abstract}

\maketitle

Quantum light-matter interfaces are key platforms in the field of quantum information. They provide storage, processing or synchronization of photonic quantum states, which can be used for applications in quantum communication, computation or sensing \cite{Bussieres2013,Sangouard2011}. One example is optical quantum memories, devices able to store and retrieve photonic quantum states. Multiplexed optical quantum memories are important in order to achieve higher data communication rates, as it is similarly done in conventional classical communications. One particularly interesting application of multiplexed quantum memories is to enhance the entanglement distribution rate in quantum repeaters \cite{Simon2007}, which in turn also facilitate their practical realization by relaxing the storage time requirements. For this application, the quantum memory should be able to store a large number of distinguishable modes and to read them out selectively. Different degrees of freedom have been considered for the multiplexed modes, such as frequency, space or time. Ensemble-based  platforms, where photonic quantum information is mapped onto collective atomic excitations, are well suited for demonstrating quantum information multiplexing. 

Cold atomic gases are currently one of the best quantum memory platforms with excellent properties demonstrated, including single photon storage and retrieval efficiency up to 90 $\%$ \cite{Bimbard2014,Yang2016,Cho2016,Vernaz-Gris2018,Wang2019} and storage time up to 220 ms \cite{Radnaev2010,Yang2016}. In particular, this system is well suited for realizing a photon pair source with embedded quantum memory following the Duan-Lukin-Cirac-Zoller protocol \cite{Duan2001}, that can be used as a quantum repeater node \cite{Chou2005,Chou2007,Chen2008}. Current multi-mode atomic memories focus mainly on spatial multiplexing, e.g. adressing modes with different wavevectors  or multiple memory cells in  different parts of the cloud \cite{Lan2009,Nicolas2014,Ding2015,Pu2017,Chrapkiewicz2017,Parniak2017,Tian2017}. Beyond spatial multiplexing, time multiplexing provides a pratical way to store multiple distinguishable modes in the same ensemble of atoms. So far, time multiplexing has been mostly studied in solid-state quantum memories based on inhomogenesously broadened rare-earth doped crystals, using the atomic frequency comb scheme \cite{Afzelius2009,Riedmatten2008,Usmani2010,Clausen2011,Saglamyurek2011,Gundogan2015,Tiranov2016,Jobez2016,
Seri2017,Kutluer2017,Laplane2017,Yang2018}. In contrast, very few experiments have investigated time multiplexing in atomic gases either by using  a controlled and reversible broadening of the spin transition \cite{Hosseini2011a,Glorieux2012, Albrecht2015, Farrera2018} or very recently by mapping photons generated in different spatial modes to different temporal modes \cite{Wen2019,Li2019}.

\begin{figure*} [ht]
	\includegraphics[width=1\textwidth]{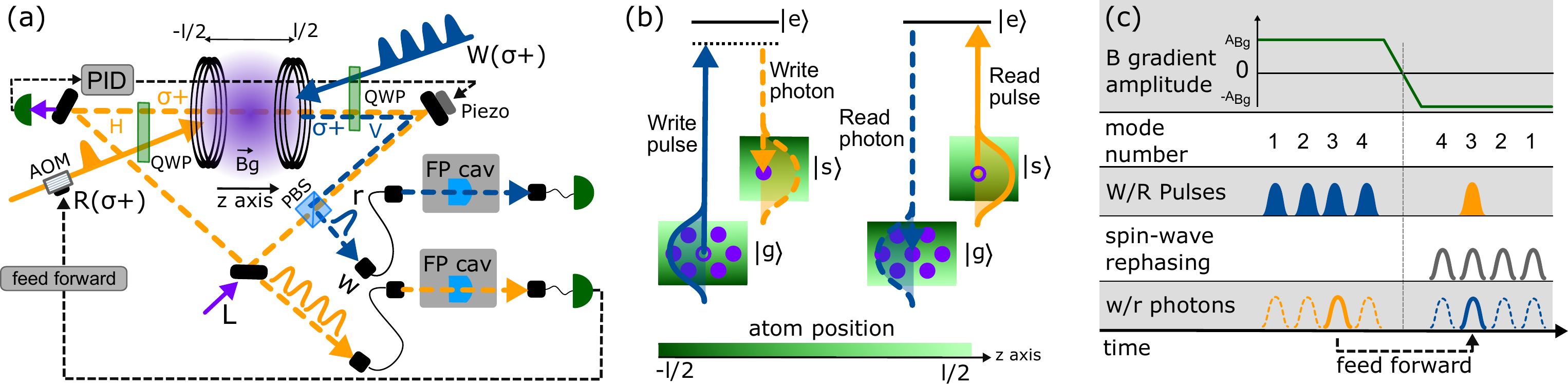}
	\caption{(a) Schematic overview of the experimental setup. W, write pulse; R, read pulse; w, write photon; r, read photon; L, cavity locking laser beam; PID, proportional-integral-derivative controller; FP cav, Fabry-Perot filtering optical cavity; l, atomic cloud length; QWP, quarter-wave plate; PBS, polarization beamsplitter. The polarizations indicated are in the atomic frame (cf. \cite{supplement}). (b) Energy levels relevant for the photon generation process. The green color gradient bars illustrate the position dependent Zeeman level energy shift along the z axis. (c) Time diagram of events occurring in the temporally multiplexed operation of the system. In this example, 4 write pulse modes are sent to the atomic cloud. A magnetic field gradient of amplitude $\rm{A_{Bg}}$ is present which is reversed to $\rm{-A_{Bg}}$ after the last write pulse mode. If a write photon is detected in the 3rd mode, a feed forward instruction sends the read pulse at the time corresponding to the 3rd-mode spin wave rephasing time.}
	\label{fig1}
\end{figure*}

Previous attempts to generate non-classically correlated pairs of photons and spin waves in multiple temporal modes in the same spatial mode have been plagued by a linear increase of the noise as a function of number of modes due to dephased spin waves \cite{Albrecht2015}. This effect prevents significant gain in photon pair generation rate, compared to the single-mode case.  In this paper, following a proposal by Simon et al \cite{Simon2010}, we demonstrate that by embedding the ensemble inside a low finesse cavity, one can substantially reduce noise from dephased spin waves. We experimentally show noise reduction by a factor 14. Subsequently, we demonstrate the generation of cavity-enhanced photons paired with spin waves in up to 10 temporal modes while preserving high quantum correlations between the photons and spin waves. This allows us to increase the spin wave - photon (photon pair) creation rate by a factor 10 (7.3), with respect to the single-mode case. The number of modes could be greatly improved by increasing the finesse of the cavity.  

In the DLCZ scheme, an off-resonant write laser pulse generates collective excitations in an atomic cloud that are correlated with Raman scattered write photons. These excitations can be mapped with high readout efficiency into read photons as long as the atomic coherence is preserved. In order to achieve temporal multiplexing, we need two additional ingredients. First, controlled dephasing and rephasing of the collective excitations that allows one to distinguish spin waves created at different temporal modes. Second, an optical cavity to reduce noise generated from the dephased excitation modes \cite{Simon2010}.

The experimental setup is shown in Fig.~\ref{fig1}a. We cool an ensemble of $\mathrm{^{87}Rb}$ atoms in a magneto-optical trap to a temperature of around $40\,\mathrm{\mu K}$. The relevant atomic levels are shown in Fig.~\ref{fig1}(b) and consist of two metastable ground states ($|g\rangle = |5^2S_{1/2},F=1,m_F=1\rangle$ and $|s\rangle = |5^2S_{1/2},F=2,m_F=1\rangle$) and one excited state ($|e\rangle = |5^2P_{3/2},F=2,m_F=2\rangle$). After optically pumping the atoms to $|g\rangle$, a write pulse with duration $\rm{\Delta t_W=266\,ns}$ drives the transition $|g\rangle\rightarrow|e\rangle$ red detuned by $\mathrm{\Delta=40MHz}$. This process probabilistically generates write photons on the $|e\rangle\rightarrow|s\rangle$ transition through spontaneous Raman scattering that are paired with collective spin excitations (atoms in $|s\rangle$). 

In order to distinguish different spin wave temporal modes, a spatial gradient magnetic field is present during writing. This causes a position dependent energy shift of the atomic levels through the Zeeman effect. The temporal evolution of the spin waves can be written as

\begin{equation}
\left|\Psi_a(t)\right\rangle=\frac{1}{\sqrt{N}}\sum_{j=1}^{N}e^{i\bold{x}_j(\bold{k}_W-\bold{k}_w)+i\int_{0}^{t}\Delta w_j(t')dt'}\left|g_1...s_j...g_N\right\rangle
\label{eqn:spinwave}
\end{equation}

where the two-photon detuning $\Delta w_j$ is different for each atom. Here, $N$ denotes the total number of atoms, $\bold{x}_j$ the initial atom position and $\bold{k}_{W(w)}$ the wavevector of the write pulse (photon). $\Delta w_j=\mu_{B}B(x_j)(g_{F(|s\rangle)}m_{F(|s\rangle)}-g_{F(|g\rangle)}m_{F(|g\rangle)})/\hbar$ where $\mu_B$ is the Bohr magneton, $B(x_j)$ is the magnetic field at the position of atom $j$, $g_{F(|s,g\rangle)}$ is the Land\'e g-factor, and $m_{F(|s,g\rangle)}$ the quantum number corresponding to the z-component of the total angular momentum. The gradient field is provided by the trapping coils.

The collective atomic excitation can be converted into a read photon by means of a read pulse resonant to the $|s\rangle\rightarrow|e\rangle$ transition. In the absence of atomic dephasing the emission will be highly efficient into a particular spatio-temporal mode thanks to collective interference of all contributing atoms. In the case of spin wave dephasing, i.e. like the one induced by the magnetic field gradient, no collective interference occurs and the read-out process will not be efficient. However, inverting the amplitude of the magnetic field gradient (and thereby inverting the phase evolution of the spin wave) eventually leads to its rephasing and efficient photon retrieval. This technique can be used to write $N_m$ different temporal modes and select a particular one to be read-out (see Fig.~\ref{fig1}c). Note that while we can trigger the phase reversal on demand, there will be a delay between this trigger and the actual read-out. This delay does not prevent however the use of our memory in a quantum repeater architecture, as explained in the supplemental material \cite{supplement}.

When reading a particular temporal mode, a major noise source arises from dephased spin waves generated in other temporal modes. During writing, spin waves are created which are paired with photons emitted into all possible modes, not only the write photon mode. Such spin waves, if in-phase, emit read photons into the corresponding phase matched (uncollected) mode and therefore not contribute to read-out noise. However, $N_m -1$ dephased spin waves will emit read photons into all directions and therefore generate noise in the read mode \cite{Simon2010, supplement}. Non-perfect rephasing of the read-out mode adds additional noise proportional to $1-p_{r|w}^{\rm{int}}$, where $p_{r|w}^{\rm{int}}$ is the intrinsic read-out efficiency. We obtain the following expression for the total probability to detect a noise photon from dephased spin waves \cite{supplement}
	
\begin{equation}
p_{r|w}^{\rm{noise}}=p(N_{\rm{m}}-p_{r|w}^{\rm{int}})\frac{\beta_r}{\beta_w}\xi_{eg}\eta_r
\label{eq:noise}.
\end{equation}

Here, $p$ is the probability to generate a spin wave - write photon pair, $\beta_{w(r)}$ is the fraction of write (read) photons that are emitted into the collected spatial mode, $\xi_{eg}$ is the branching ratio corresponding to the $|e\rangle-|g\rangle$ transition, and $\eta_r$ is the detection efficiency of the read photons. In order to decrease this noise one can increase the ratio $\beta_w$ of excitations paired with write photons over excitations paired with photons emitted into other spatial modes. This is achieved with an optical cavity enhancing the photon emission into the write photon spatial mode. Such a cavity is schematically described in Fig.~\ref{fig1}a. In order to not simultaneously increase $\beta_r$ while increasing $\beta_w$, the read photon has orthogonal polarization from the cavity mode and is decoupled from the cavity by a PBS.

%\begin{figure}
%	\includegraphics[width=.48\textwidth]{fig2.pdf}
%	\includegraphics[width=.48\textwidth]{fig3.pdf}
%	\caption{(a) Write photon emission probability as a function of the cavity resonance frequency. Frequency zero corresponds to the center of the $|e\rangle-|s\rangle$ transition. The green solid line represents the emission without cavity enhancement. (b) Read photon detection probability from dephased spin waves as a function of the spin wave excitation probability. The spin wave is read out after $\rm{1.2 \mu s}$ of storage time. This time is much longer than the spin wave dephasing time set by the B field gradient \cite{Albrecht2015}. Blue (green) data is taken with (without) cavity enhancement.}
%	\label{fig23}
%\end{figure}

%\begin{figure}
%	\begin{subfigure}{0.45\linewidth}
%		\includegraphics[width=\linewidth]{fig2.pdf}
%	\end{subfigure}
%	\begin{subfigure}{0.45\linewidth}
%		\includegraphics[width=\linewidth]{fig3.pdf}
%	\end{subfigure}
%	\caption{(a) Write photon emission probability as a function of the cavity resonance frequency. Frequency zero corresponds to the center of the $|e\rangle-|s\rangle$ transition. The green solid line represents the emission without cavity enhancement. (b) Read photon detection probability from dephased spin waves as a function of the spin wave excitation probability. The spin wave is read out after $\rm{1.2 \mu s}$ of storage time. This time is much longer than the spin wave dephasing time set by the B field gradient \cite{Albrecht2015}. Blue (green) data is taken with (without) cavity enhancement.}
%	\label{fig23}
%\end{figure}

\begin{figure}
	\begin{minipage}{0.49\linewidth}
		\includegraphics[width=\linewidth]{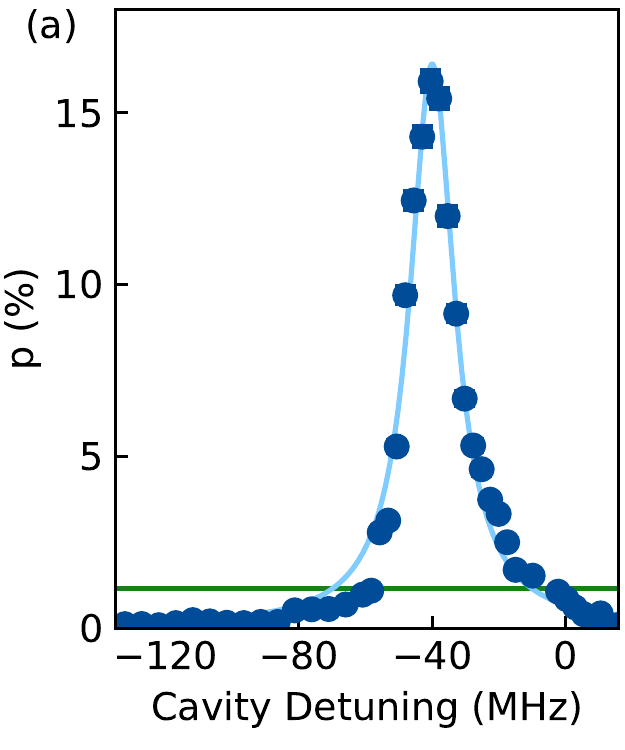}
	\end{minipage}\hfill
	\begin{minipage}{0.49\linewidth}
		\includegraphics[width=\linewidth]{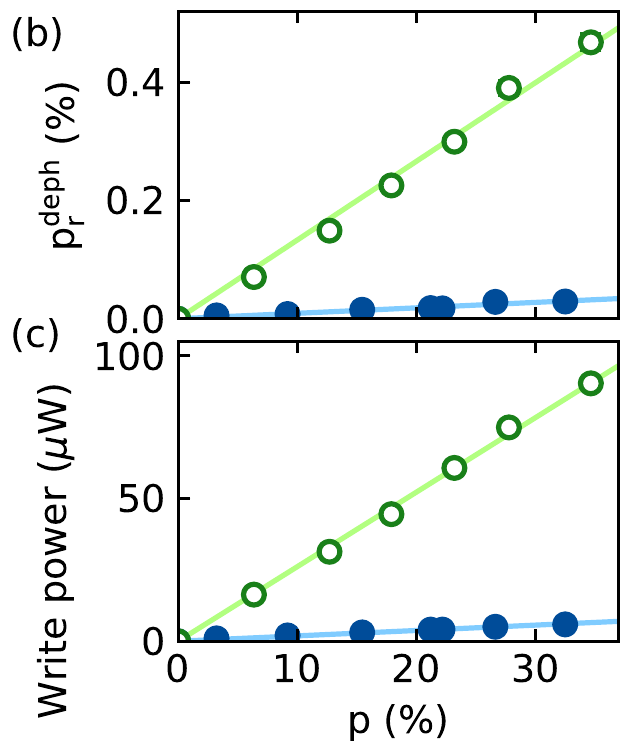}
	\end{minipage}
	\caption{(a) Write photon emission probability as a function of the cavity resonance frequency. Frequency zero corresponds to the center of the $|e\rangle-|s\rangle$ transition. The green solid line represents the emission without cavity enhancement. (b) Read photon detection probability from dephased spin waves and (c) write pulse power as a function of the spin wave excitation probability. The spin wave is read out after $\rm{1.2 \mu s}$ of storage time. This time is much longer than the spin wave dephasing time set by the B field gradient \cite{Albrecht2015}. Blue (green) data is taken with (without) cavity enhancement.}
	\label{fig23}
\end{figure}

Fig.~\ref{fig23}a characterizes the cavity enhanced write photon emission. The cavity resonance frequency is changed by moving one of the cavity mirrors with a piezo-electric device. When the cavity resonance matches the write photon transition, photon emission is enhanced. However, when the two frequencies differ by more than the cavity linewidth, the emission is suppressed. At resonance we observe enhancement of $p_{\rm enh}^{\rm c}/p=14.3(6)$. Here, $p^{c} \, (p)$ is the write photon emission probability with (without) cavity. This is close to the expected value of $2F/\pi$ \cite{Wilmsen1999},  while out of resonance inhibition is $p_{\rm inh}^{c}/p=0.078(3)$. The spectral width of the emission is $\rm 16.6 MHz$ and matches the cavity linewidth (for more details on the cavity parameters see \cite{supplement}).
Note that the effective enhancement of the write photon detection probability with cavity is reduced by the cavity escape efficiency, which for our implementation is 56\%.

As mentioned before, the cavity enhancement of the write process allows for suppression of the read photon noise generated from dephased excitations. This is quantified in Fig.~\ref{fig23}b and~\ref{fig23}c. In order to measure this dephased noise, a magnetic field gradient is applied during writing without field inversion before read-out. This causes a rapid dephasing of the generated spin waves and hence all the read-out photons are generated through interaction of the read pulse with dephased spin waves. In Fig.~\ref{fig23}b (\ref{fig23}c), the read photon detection probability (write pulse power) is shown as a function of the write photon generation probability $p$. We observe that for the same excitation probability, the noise (write pulse power) is 14.4(7) (13.9(3)) times lower in the cavity enhanced situation (which is compatible with the cavity enhancement observed in Fig.~\ref{fig23}a).  The enhancement gives an approximate upper bound on the number of modes that can be used in a temporally multiplexed operation of the system.

\begin{figure}
	\includegraphics[width=.48\textwidth]{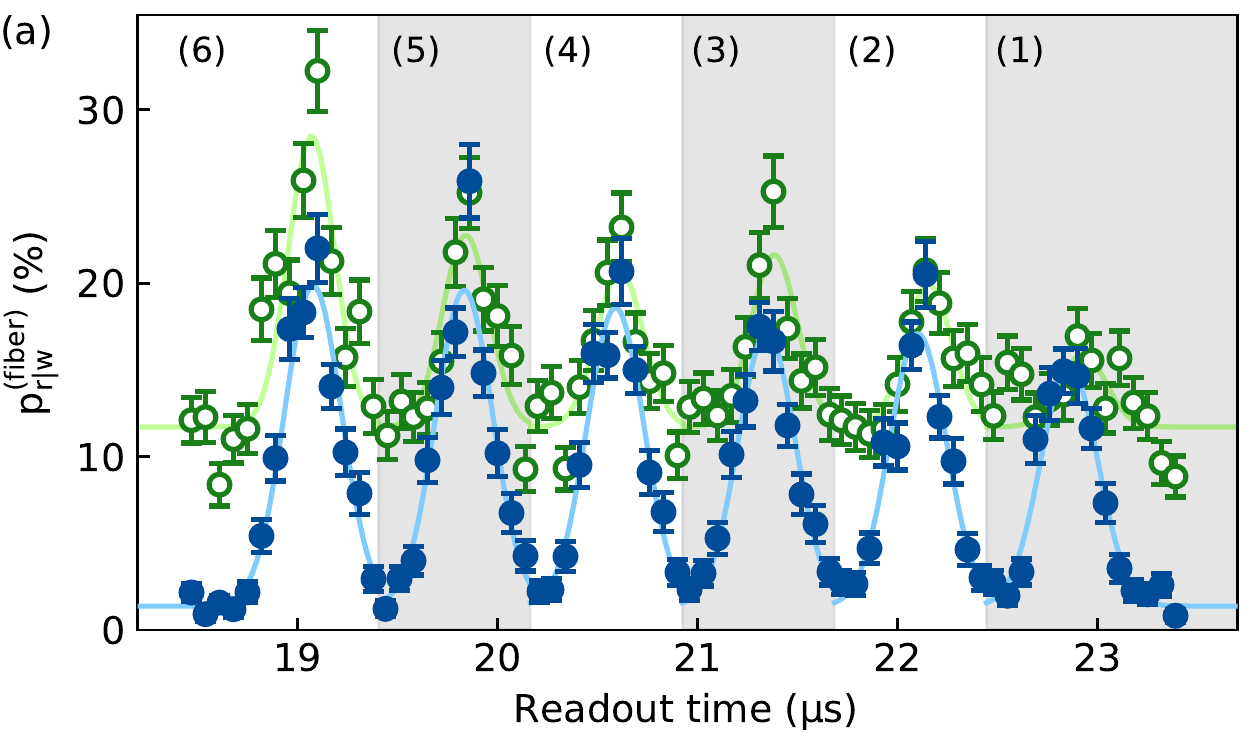}
	\includegraphics[width=.48\textwidth]{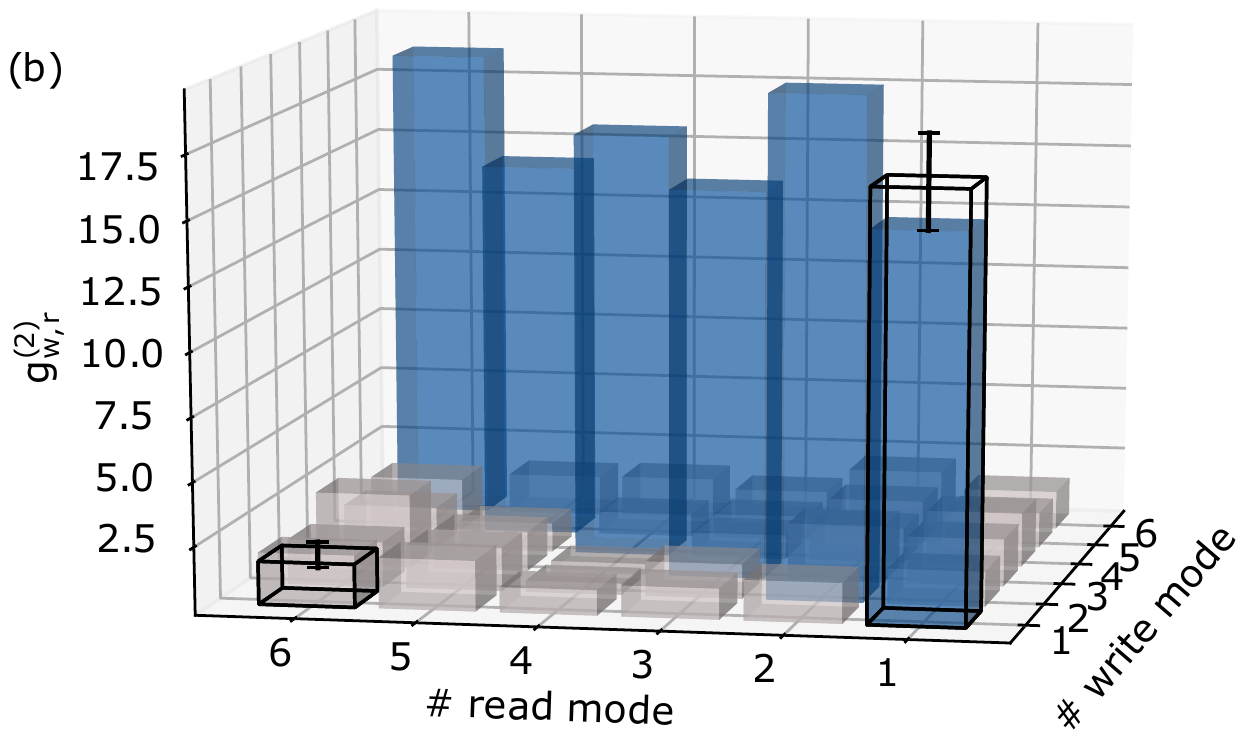}
	\caption{(a) Probability to collect an heralded read photon into the read fiber as a function of the read-out time for 6 different temporal modes. Time zero corresponds to time of writing of the last write mode (6). With cavity enhancement, intrinsic retrieval efficiency for the first mode is $p^{int}_{r|w}\approx26\%$.  Blue (green) data is taken with (without) cavity enhancement. For both, single-mode excitation probability is $p_{1m} \approx 0.04$. Solid lines are a Gaussian fit of each retrieval peak. (b) Individual cross-correlation function between the different 6 write and read modes with cavity. The two bars in solid black lines at positions (1,1) and (1,6) represent the average for the diagonal and the off-diagonal values, respectively. }
	\label{fig45}
\end{figure}

After characterizing cavity-enhanced emission and noise reduction, we now compare temporally multiplexed storage with and without enhancement, as depicted in Fig.~\ref{fig1}c. Fig.~\ref{fig45} shows a situation in which 6 write pulse modes are sent to the atomic cloud. In Fig.~\ref{fig45}a, after the 6-modes write process, the magnetic field gradient is reversed. Upon detection of a write photon, we use a feed forward instruction in order to scan the read-out around the expected rephasing time. We observe 6 peaks corresponding to the rephasing of each of the 6 spin wave modes. This figure shows 6 different data sets (separated with white and grey backgrounds) corresponding to write photon detection in different temporal modes. The ratio between the SNR achieved with/without cavity enhancement of $\approx 14$
highlights the noise reduction achieved with cavity. For the cavity case, Fig.~\ref{fig45}b characterizes the cross-correlation function between the write and read photons (defined as $g^{(2)}_{w,r}=p_{w,r}/(p_wp_r)$, where $p_{w,r}$ is the probability to detect a coincidence between write and read photon, and $p_w  (p_r)$ is the probability to detect a write (read) photon) in all the 36 possible combinations of six write and six read modes. The correlations are preserved when the read mode corresponds to the write mode (weighted average 16.6(1.8)). However, little crosstalk is observed when the read mode is different from the considered write mode (weighted average 1.7(0.5)).

\begin{figure}
	\includegraphics[width=.48\textwidth]{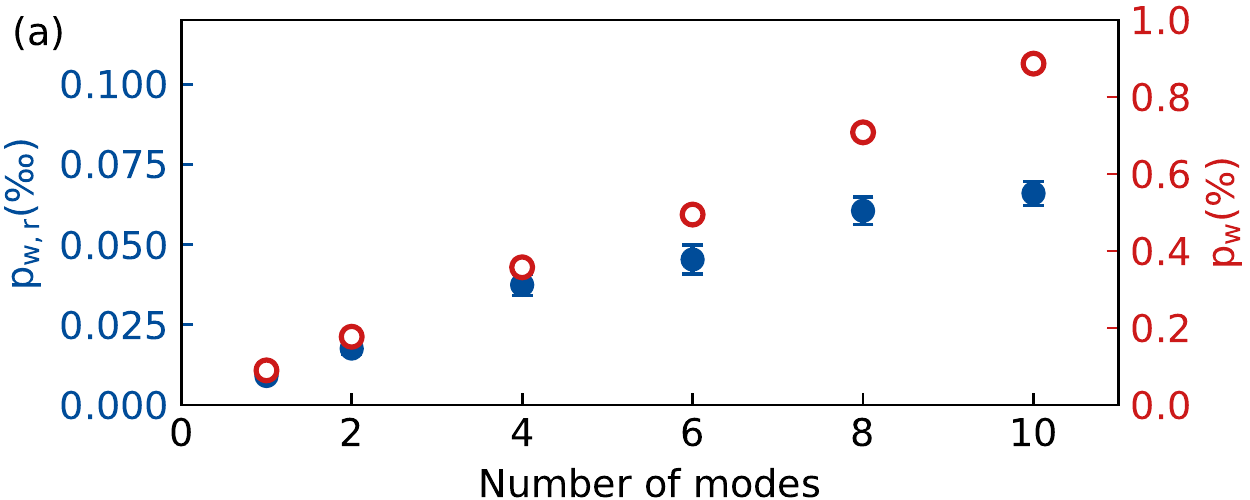}
	\includegraphics[width=.48\textwidth]{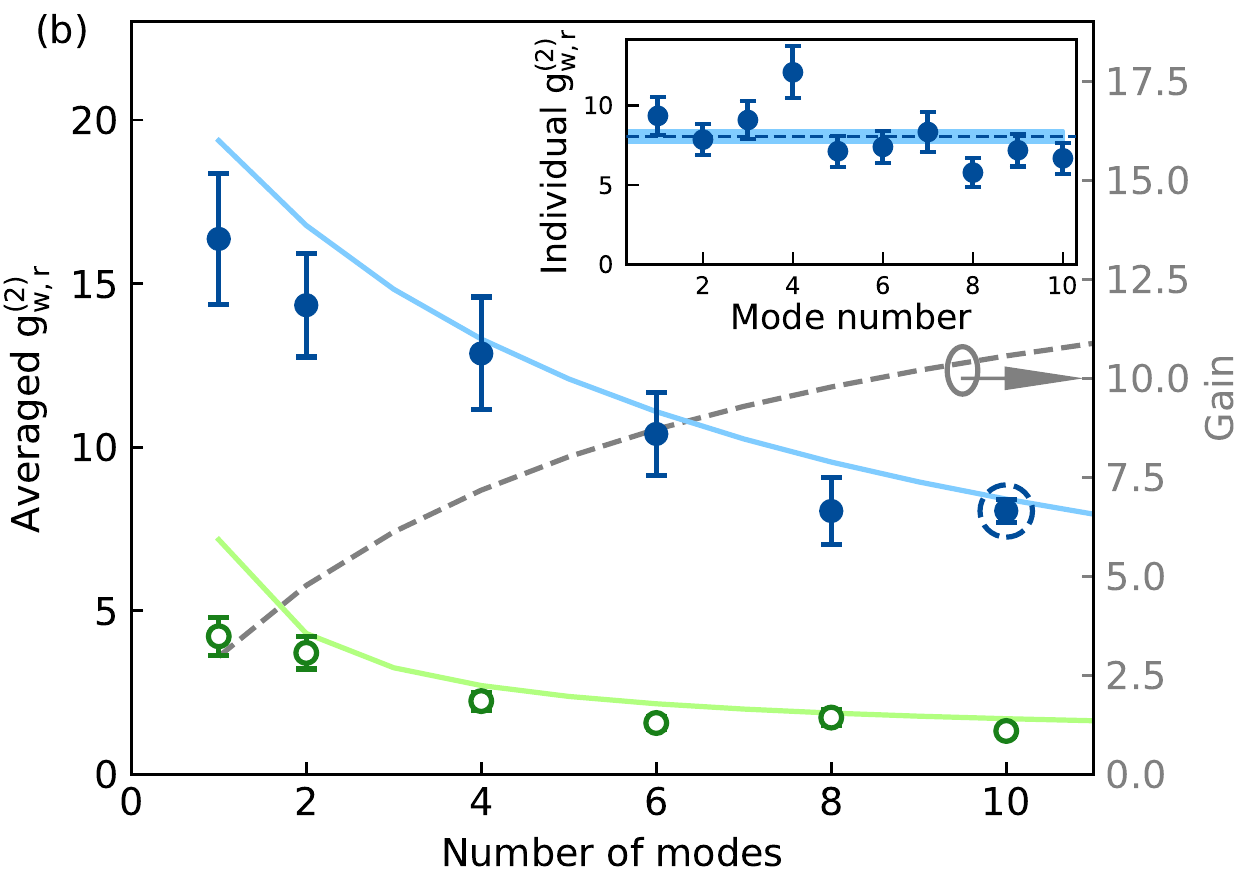}	
	\caption{(a) Write and total write-read detection probability as a function of the number of temporal modes with cavity. (b) Averaged correlation function between write and read photons as a function of the number of modes. Average is computed based on the sum of coincidence and noise counts from all modes. Error bar is one stdev, again based on the sum of counts in all modes. Blue (green) data is taken with (without) cavity enhancement. Grey dashed line shows the gain $\rm (g^{(2),\,c}_{w,r} -1)/(g^{(2)}_{w,r} -1) $ in cross-correlation enabled by the cavity, as a function of the number of modes. Here, $g^{(2),\,c}_{w,r}$ ($g^{(2)}_{w,r}$) is the cross-correlation value with (without) cavity. Inset shows $g^{(2)}_{w,r}$ of each mode for the 10-mode data point. Single-mode excitation probability is $p_{1m} \approx 0.045$ for all measurements.}
	\label{fig67}
\end{figure}

Finally, we characterize the cavity enhanced temporally multiplexed operation of the system. After a train of $N_m$ write pulses and the recording of a write photon, the magnetic field gradient is inverted. A read pulse is sent at the expected rephasing time of its paired spin excitation. In Fig.~\ref{fig67}a, scanning the number of modes, we observe that the write photon detection probability per write pulse train, and hence the probability to create a spin wave - photon pair, increases linearly with $N_m$, while the write-read photon coincidence detection probability has a slightly worse scaling. This can be explained by the reduced readout efficiency as a function of storage time and by magnetic field fluctuations (cf. \cite{supplement}). Nevertheless, for 10 modes we obtain a total rate enhancement of 7.3. In Fig.~\ref{fig67}b, again scanning $N_m$, we show the averaged value of $g^{(2)}_{w,r}$ across $N_m$ modes. We notice that the multiplexed operation has a much stronger degradation impact on the correlation between the write and read photons when no cavity is present. The cavity significantly reduces the impact of the dephased spin waves on the quality of the correlations. It is also remarkable that $g^{(2)}_{w,r}$ is higher with cavity enhancement in the case of just 1 temporal mode. This highlights the imperfect rephasing of the spin wave, leading to dephased noise that is suppressed by the cavity. This is predicted by Eq.~\ref{eq:noise} for read out efficiencies $<$1 and explained in more detail in \cite{supplement}. Moreover, the values $g^{(2)}_{w,r}>2$ are an evidence of quantum correlation between the write and read photons, assuming thermal statistics for the individual write and read modes. For 10 modes, we also measured the  averaged heralded autocorrelation of the generated single photon and found $g^{(2)}_{r,r|w}$=0.36(0.25)$<$ 1, confirming the non-classical nature of the emitted photons. 

The maximal number of temporal modes is currently limited by the finesse of our cavity, which is in turn limited by the optical intra-cavity loss, mostly given by the windows of our vacuum chamber. This loss is also responsible for the low escape efficiency in our current experiment (cf. \cite{supplement}). This is however not a fundamental limitation. By implementing a cavity inside the vacuum chamber, a much higher cavity finesse could be achieved while keeping a high escape efficiency, such that $N_m$ $>$ 100 should be readily possible. For such a large number of modes, the next limitation is the spin wave storage time. With write modes separated by 800 ns as in our implementation, memory lifetimes of $2*80\mu s$ become necessary. However, DLCZ experiments with cold atoms in optical lattices have shown much longer storage times of up to 200 ms \cite{Radnaev2010, Yang2016}. Reaching long storage times is facilitated by the use of magnetically insensitive transitions to minimize decoherence by magnetic fluctuations. These transitions are not directly compatible with the broadening using magnetic gradients as demonstrated in our current proof of principle experiment. However, several solutions could be applied, e.g. transferring the excitations to clock transitions after the write pulse train \cite{Jiang2016} or using light shifts for inducing and reversing the broadening \cite{Sparkes2010,Parniak2019}. Finally, we note that the gain in coincidence count rate due to the multi-mode operation is only present for a fixed repetition rate of the experiment. This is for example the case for quantum repeater applications, where entanglement between distant quantum memories must be heralded. In that situation, the repetition rate of the entanglement attempts is limited to $R=c/L_0$ where $L_0$ is the distance between the ensembles. For example, for $L$=100 km, $R$=2 kHz. In that case, temporal multiplexing would increase the entanglement rate by a factor $N_m$ for low success probability \cite{Simon2007}.

In conclusion, we presented a temporally multiplexed quantum repeater node based on cold atomic ensembles. By implementing a controlled inhomogeneous broadening of the spin transition, we generated distinguishable spin waves. We significantly reduced noise due to dephased spin waves by embedding the ensemble inside a low finesse optical cavity. This allowed us to demonstrate multiplexed generation of non-classical spin wave - photon pairs in up to 10 temporal modes, enabling a corresponding increase in generation rate. These correlated pairs could also serve as a source of high-dimensional light-matter entanglement in time. The multiplexing capability can be further enhanced by using a higher finesse cavity or by combining temporal multiplexing with other techniques such as frequency or spatial multiplexing. 

\begin{acknowledgments} We acknowledge support by the Spanish Ministry of Economy and Competitiveness (MINECO) and the Fondo Europeo de Desarrollo Regional (FEDER) through grant FIS2015-69535-R, by MINECO Severo Ochoa through grant SEV-2015-0522, by the Gordon and Betty Moore Foundation through Grant
No. GBMF7446 to H. d. R, .by Fundaci\'{o}  Privada Cellex and by the CERCA programme of the Generalitat de Catalunya. P.F.  acknowledges financial support by the Cellex ICFO MPQ
research fellowship program. L.H. acknowledges funding from the European Union’s Horizon 2020 research and innovation programme under the Marie Skłodowska-Curie grant agreement No 713729.\\
\end{acknowledgments}

\end{document}

\title{Supplemental Material for: \\
	A cold atom temporally multiplexed quantum memory with cavity-enhanced
	noise suppression}%

\author{Lukas Heller}
\email[Contact: ]{lukas.heller@icfo.eu}
\affiliation{ICFO-Institut de Ciencies Fotoniques, The Barcelona Institute of Science and Technology, 08860 Castelldefels (Barcelona), Spain}

\author{Pau Farrera}
\email[Contact: ]{pau.farrera@icfo.eu}
\altaffiliation{Present address: Max-Planck-Institut für Quantenoptik,  Hans-Kopfermann-Strasse  1,  85748  Garching,  Germany}
\affiliation{ICFO-Institut de Ciencies Fotoniques, The Barcelona Institute of Science and Technology, 08860 Castelldefels (Barcelona), Spain}

\author{Georg Heinze}
\altaffiliation{Present address: TOPTICA Projects GmbH, Lochhamer Schlag 19, 82166 Gräfelfing, Germany}
\affiliation{ICFO-Institut de Ciencies Fotoniques, The Barcelona Institute of Science and Technology, 08860 Castelldefels (Barcelona), Spain}

\author{Hugues de Riedmatten}
\homepage{http://qpsa.icfo.es}
\affiliation{ICFO-Institut de Ciencies Fotoniques, The Barcelona Institute of Science and Technology, 08860 Castelldefels (Barcelona), Spain}
\affiliation{ICREA-Instituci\'{o} Catalana de Recerca i Estudis Avan\c cats, 08015 Barcelona, Spain}%

\begin{abstract}
In this document, we give more details about theoretical concepts and technical issues of our experiment. In particular, we characterize the impact of the write pulse duration on cavity enhancement and readout efficiency, describe theoretically the cavity enhancement and simulate the cross-correlation $g^{(2)}_{w,r}$ for different experimental conditions. Subsequently, we characterize the effect of cavity enhancement in single-mode operation, argue the choice of cavity parameters and detail the transition scheme used in this experiment. Finally, we also discuss the deviation from linear increase of coincidence rate with number of modes, as seen in Fig. 4(a) of the main paper.
\end{abstract}

\maketitle

\section{Write pulse duration characterization}

Two important figures of merit are closely linked to the write pulse duration: the cavity enhancement and the readout efficiency. As mentioned in the main text, the write pulse duration is set to a full-width at half-maximum (FWHM) of $\rm{\Delta t_W=266\,ns}$. This duration is chosen considering the above figures of merit, as explained in this section.

\begin{figure}[ht]
	\includegraphics[width=.48\textwidth]{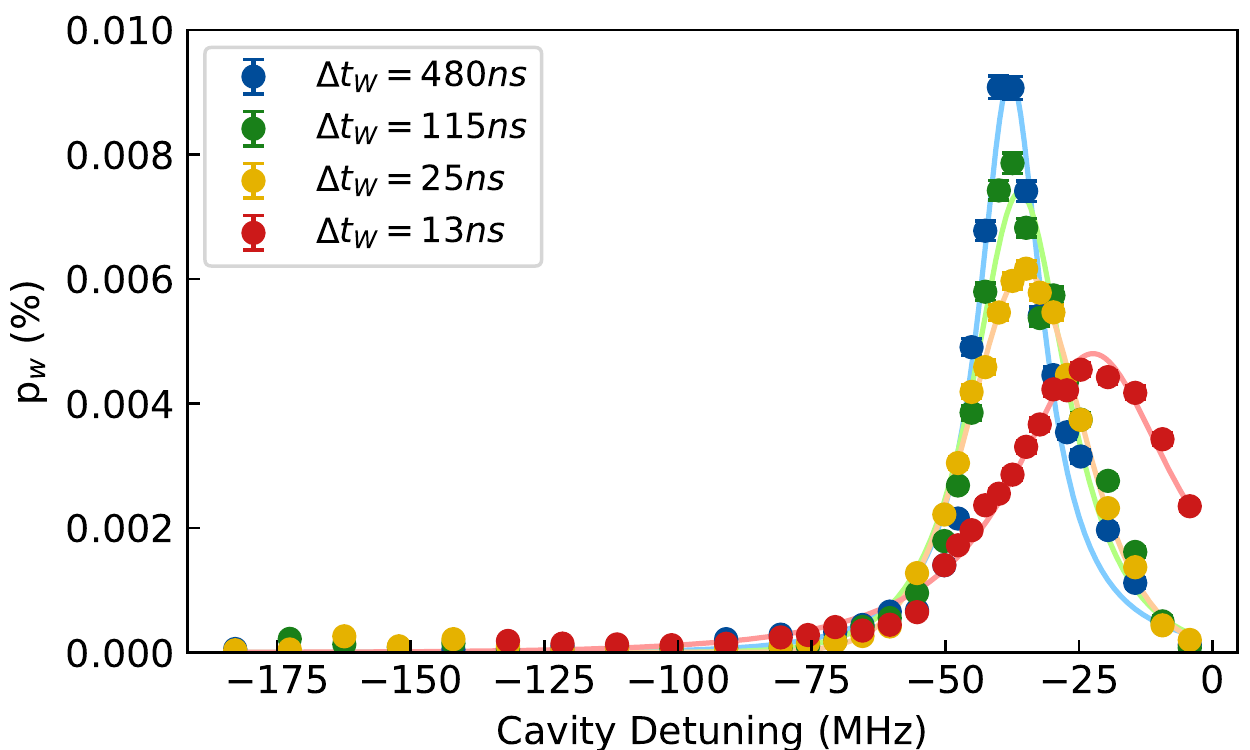}
	\caption{Write photon detection probability as a function of the cavity resonance frequency (scanned by changing the cavity piezo voltage). The different colors represent different durations $\Delta t_W$ (FWHM) of the write pulse. The zero point in the cavity detuning corresponds to the center of the $|s\rangle-|e\rangle$ transition. The tendency for shorter pulses to shift towards zero detuning is attributed to increasing frequency overlap (and therefore increasing interaction) of the write pulse with the $|s\rangle-|e\rangle$ transition. Write pulse energy is constant for all traces.}
	\label{fig1}
\end{figure}

\paragraph{Cavity Enhancement}
The write pulse temporal duration is inversely proportional to its frequency spectrum. However, only the part of the write pulse spectrum that overlaps with the cavity transmission spectrum can be effectively enhanced. This effect is shown in Fig.~\ref{fig1} where we compare the write photon detection probability $p_{w}$ for different pulse lengths, scanning the cavity detuning. For long (narrow-band) write pulses, the cavity enhancement spectrum is given by the cavity transmission spectrum. However, when decreasing the write pulse duration, its spectrum broadens. In this case the cavity transmission and write pulse spectra get convoluted, and the cavity enhancement spectrum becomes broader, showing a lower maximum enhancement. From this measurement we infer that the write pulse has to be longer than $\approx 100$ ns in order to obtain the maximum cavity enhancement for the current cavity setup.

\begin{figure}[ht]
	\includegraphics[width=.48\textwidth]{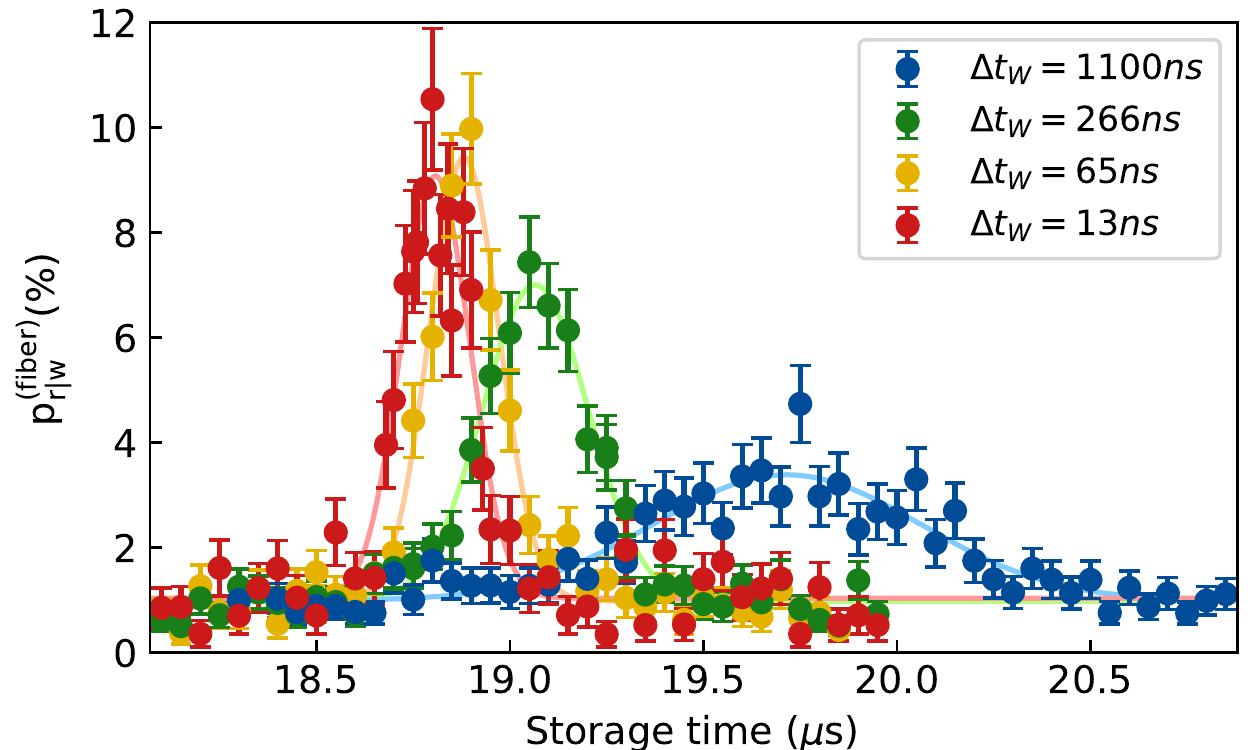}
	\caption{Probability to collect an heralded read photon into the read fiber as a function as a function of storage time. The different colors represent different durations of the write pulse, which have an impact on the width and amplitude of the rephasing peak. The peaks are slightly shifted in time for technical reasons. Maximum efficiency for $\Delta t_W = 266$ns is slightly lower than in the main text because experimental conditions were not yet optimum.}
	\label{fig2}
\end{figure}

\paragraph{Readout efficiency}
The write pulse duration also affects the controlled rephasing echo profile. When the write pulse temporal duration is comparable to or longer than the spin wave rephasing profile (which is given by the applied inhomogeneous broadening), excitations generated at different times within the same write pulse will rephase at different times. As opposed to the DLCZ protocol in a homogeneously broadened medium, it is not possible anymore to retrieve the entire spin wave at a specific retrieval time. Therefore, the retrieval efficiency $p_{r|w} = p_{w,r}/p_w$, considering all the possible spin wave creation times, will decrease. This can be seen in Fig.~\ref{fig2} where we compare the echo profile for different write pulse lengths. From this measurement we infer that the write pulse has to be on the order of $\approx 100$ ns in order to obtain decent retrieval efficiency for the current inhomogeneous broadening.

While cavity enhancement is the highest for longer write photons, the spin-wave echo shows a higher retrieval efficiency for shorter write pulses. The write pulse that we choose is a trade-off between enhancement and efficiency. While this would indicate $\rm{\Delta t_W \approx 100\,ns}$, we use a slightly longer write pulse of $\rm{\Delta t_W=266\,ns}$ as longer rephasing peaks are less affected by fluctuations in rephasing time (see below).

\section{Theory model for the cavity enhancement}

As explained in the supplemental material of \cite{Farrera2016a}, when describing theoretically the write and read photon detection probabilities in the presence of spin wave dephasing, one has to consider both the coherent and the incoherent emission processes. This can be described by the following equations: 

\begin{eqnarray}
p_{w}&=&p\eta_{w} \label{eq:1}\\
p_r &=&pp_{r|w}^{int}\eta_{r}+N_s[1-p_{r|w}^{int}]\beta_r\xi_{eg}\eta_r \label{eq:2}\\
p_{w,r} &=&pp_{r|w}^{int}\eta_{w}\eta_{r}+p\eta_{w}N_s[1-p_{r|w}^{int}]\beta_r\xi_{eg}\eta_r \label{eq:3}
\end{eqnarray}

where $p$ is the spin wave excitation probability, $\eta_{w(r)}$ is the write (read) photon total detection efficiency, $p_{r|w}^{int}$ is the intrinsic readout efficiency, $\beta_{w(r)}$ is the fraction of solid angle corresponding to the write (read) photon collection mode, $N_s=p/\beta_{w}$, is the total number of created spin excitations, and $\xi_{eg}$ is the branching ratio corresponding to the read photon transition.

Let's now consider the situation with multiple writing temporal modes and subsequent readout of a specific mode at the rephasing time of the corresponding spin wave. The read photon emission contributions from the rephased spin wave can be described by Eq.~\ref{eq:1}-\ref{eq:3}. However, in this situation one also has to consider the incoherent read photon emission from spin wave modes that are not rephased. The photon contributions to $p_r$ and $p_{w,r}$ from these dephased modes is:

\begin{eqnarray}
p_r^{deph}&=&N_s^{deph}\beta_r\xi_{eg}\eta_r \label{eq:4}\\
p_{w,r}^{deph}&=&p\eta_wN_s^{deph}\beta_r\xi_{eg}\eta_r \label{eq:5}
\end{eqnarray}

where $N_s^{deph}=p(N_{m}-1)/\beta_w$ is the total number of spin excitations generated by write pulses corresponding to modes different from the rephased one. Adding the contributions~\ref{eq:4}-\ref{eq:5} to~\ref{eq:2}-\ref{eq:3}, we can write the expressions for the detected read-out efficiency ($p_{r|w}=p_{r,w}/p_w$) and the cross correlation function ($g^{(2)}_{w,r}=p_{w,r}/(p_wp_r)$)as

\begin{eqnarray} 
 p_{r|w}&=&p_{r|w}^{int}\eta_{r}+p(N_{\rm{m}}-p_{r|w}^{\rm{int}})\frac{\beta_r}{\beta_w}\xi_{eg}\eta_r \label{eq:6}\\
 g^{(2)}_{w,r}&=&1+\frac{p_{r|w}^{int}(1-p)}{pp_{r|w}^{int}+p(N_{\rm{m}}-p_{r|w}^{\rm{int}})\frac{\beta_r}{\beta_w}\xi_{eg}}. \label{eq:7}
\end{eqnarray}

We can observe that increasing $N_{\rm m}$ increases the incoherent therm in $p_{r|w}$ and decreases $g^{(2)}_{w,r}$. However this can be overcome by increasing the ratio $\beta_w/\beta_r$. In our experiment we obtain this by having an optical cavity resonant with the write photon but decoupled from the read photon mode. In such a situation the ratio $\beta_w/\beta_r$ is equal to the write photon cavity enhancement factor $\frac{2F}{\pi}$ (compare also simulations in  Fig.~\ref{fig3}-\ref{fig4}).

Please note that experiments with cavities for both write and read process have been realized as well \cite{Bao2012}, with the aim of improving the read-out efficiency. However, in this case the ratio $\beta_w/\beta_r$ is unity, making temporally multimode operation impossible. For achieving higher retrieval efficiencies in the current setup, the optical depth needs to be improved by other means, for example by compressing the atoms during the MOT phase.
 
\section{Multi-mode Cross-correlation Simulations -- Dependence on cavity enhancement and retrieval efficiency}
 
For the following simulations, the model from the previous section is used. No dependence of retrieval efficiency on storage time was considered, i.e. assuming $N_{m}\tau_{m} \ll \tau_{mem}$. Here, $\tau_{m}$ is the single-mode duration and $\tau_{mem}$ the memory lifetime.
 
\begin{figure}[ht]
 	\includegraphics[width=.48\textwidth]{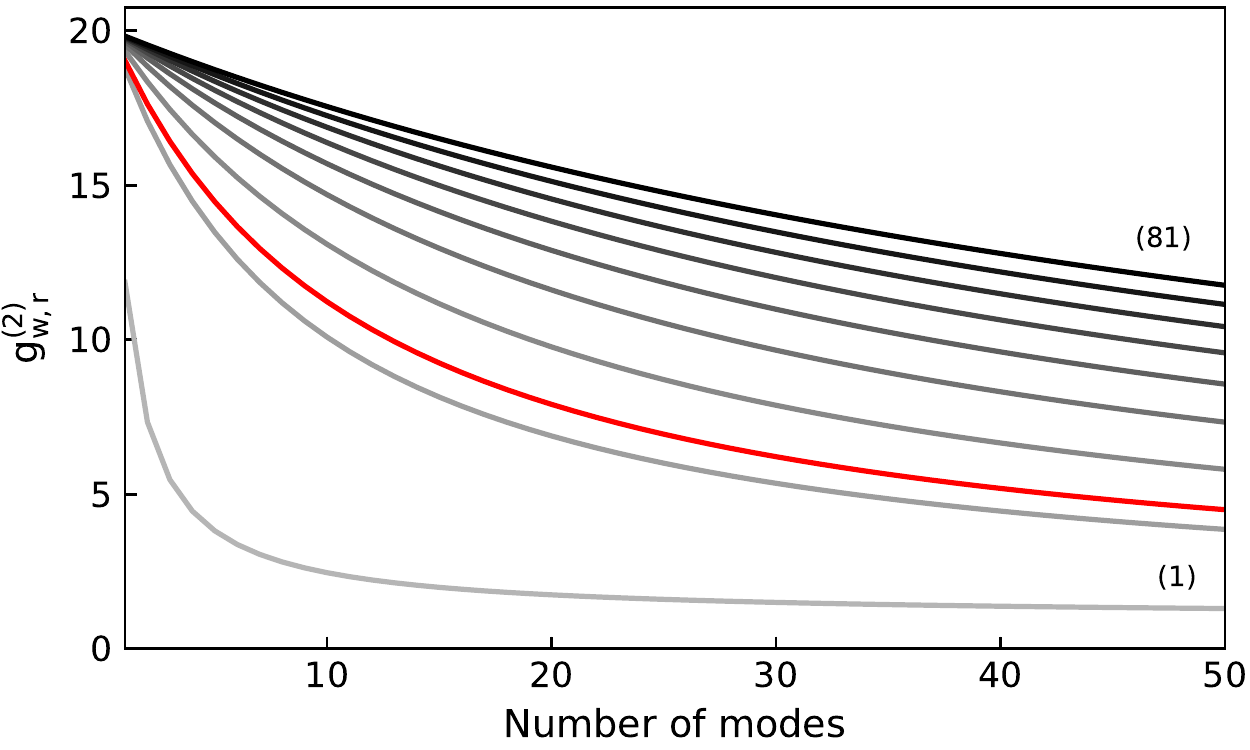}
 	\caption{$g^{(2)}_{w,r}$ as a function of number of input modes, for different values of the cavity enhancement between 1 (no cavity) and 81, in steps of 10. In red the current setup with a cavity enhancement of 14.}
 	\label{fig3}
\end{figure}

For the given memory setup and parameters, Fig.~\ref{fig3} shows the scaling of cross-correlation as a function of mode number, plotted for different cavity enhancements. In red we show the current setup performance. As apparent from the simulation, for the multi-mode case substantial improvement is still possible by increasing the cavity's finesse, However, for the single-mode case only small improvement is expected.
 
\begin{figure}[ht]
	\includegraphics[width=.48\textwidth]{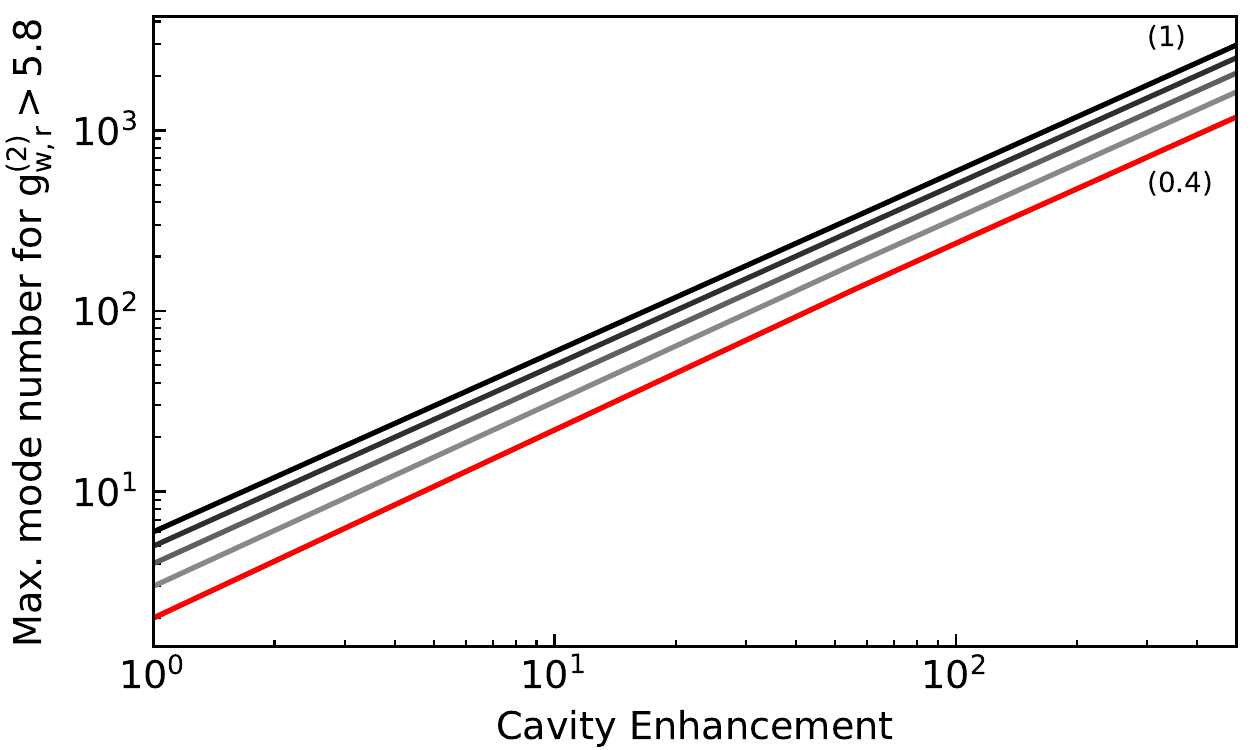}
	\caption{Maximum number of input modes to comply with {$\rm g^{(2)}_{w,r} > 5.8$} vs. cavity enhancement, for different values of the retrieval efficiency between 0.4 (current setup) and unity, in steps of 0.15. In red the current setup with a retrieval efficiency of 0.4.}
	\label{fig4}
\end{figure}

We observe that adding modes to the memory reduces quantum correlations due to increased noise. Given a minimum threshold for the cross-corellation function, it is instructive to investigate up to how many modes the memory supports, still complying with this threshold. As a threshold we choose a value of 5.8 (minimum value necessary for violating Bell's inequality \cite{Riedmatten2006}). For a given enhancement, Fig.~\ref{fig4} shows this maximum number as a function of cavity enhancement, plotted for different retrieval efficiencies. We find a linear behaviour on the cavity enhancement. Furthermore, we see that great improvement is still possible by increasing the retrieval efficiency of the memory towards unity.

\section{Effect of cavity enhancement in single-mode operation}
 
\begin{figure}[ht]
	\includegraphics[width=.48\textwidth]{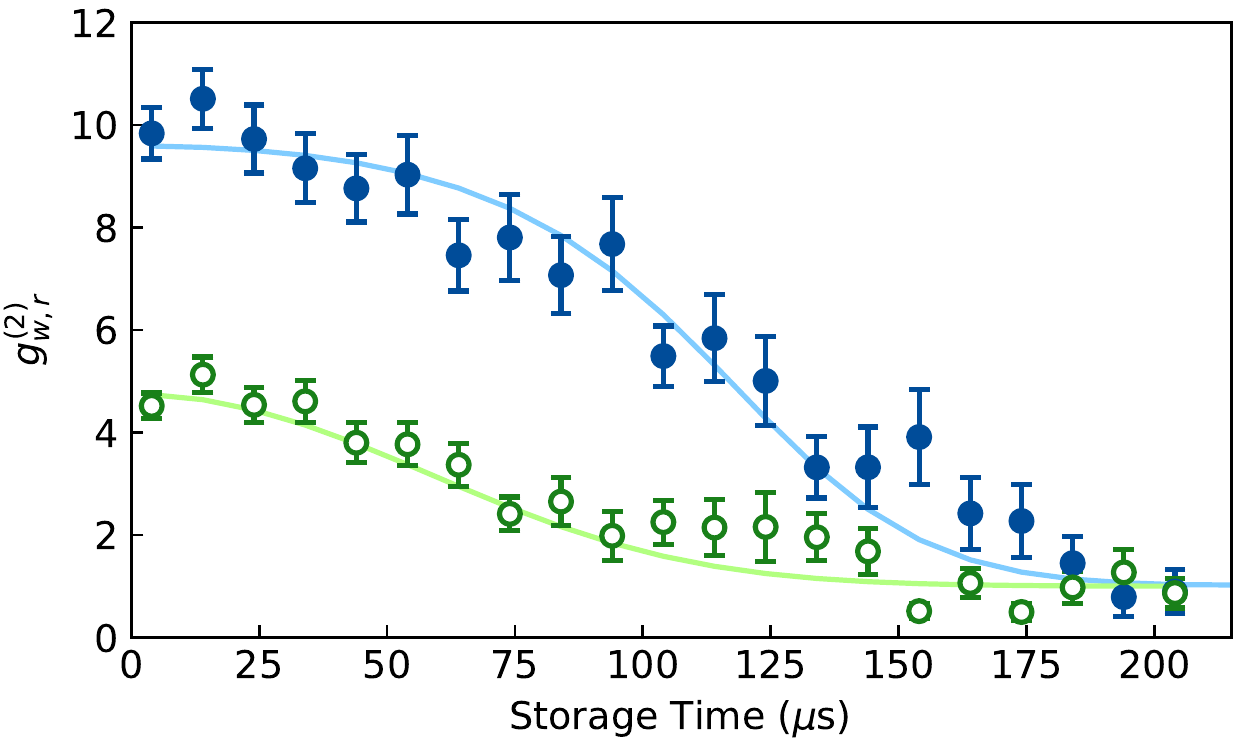}
	\caption{Cross-correlation function $g^{(2)}_{w,r}$ as a function of storage time for DLCZ storage protocol in homogeneously broadened medium. Blue (green) data is taken with (without) cavity enhancement. $p \approx 10\%$ for both traces. Solid lines simulate expected behaviour, assuming a cavity enhancement of 14 for cavity enhanced emission and a 1/$e$ memory lifetime of $\tau \approx 72 \mu s$ in both cases.}
	\label{fig5}
\end{figure}

Up to now, the cavity was employed solely during multi-mode operation. Now, we investigate the performance of the memory in single-mode operation, i.e. for the DLCZ storage protocol in the homogeneously broadened medium. In Fig.~\ref{fig5} we show measured cross-correlation as a function of storage time. For the same excitation probability $p$, we compare the performance with and without cavity enhancement of the write photon generation process. In both cases, measured data is in good agreement with the expected behaviour (solid lines, computed from Eq.~\ref{eq:7} with $p$, memory lifetime $\tau$ and cavity enhancement as stated in the figure caption). 
 
Two important findings can be deduced from this measurement. Firstly, already at time zero the performance of the cavity enhanced memory is superior to the one without enhancement. This is a consequence of the non-unity retrieval efficiency of our memory. The retrieval efficiency at zero storage time is limited by insufficient optical depth and re-absorption (motional or spurious magnetic dephasing mechanisms do not yet play a role, compare also Fig.~\ref{fig3} and Fig.~\ref{fig4}). The gain $\rm (g^{(2),\,c}_{w,r} -1)/(g^{(2)}_{w,r} -1)$, where $g^{(2),\,c}_{w,r}$ ($g^{(2)}_{w,r}$) is the cross-correlation function with (without) cavity, at zero storage time is 2.3. Secondly, the correlation decay versus storage time is much more apparent when the cavity is not employed. With cavity, $\rm (g^{(2),\,c}_{w,r} -1)$ only drops by 15\% between time zero and the 1/$e$ memory lifetime $\tau \approx 72 \mu s$. However, the same figure drops by 60\% without cavity. This shows the noise reduction capability of the cavity and can be further enhanced by enhancing the finesse.
 
These results demonstrate that cavity enhanced write emission is not only beneficial in multi-mode operation (as investigated in the main paper) but can also greatly improve the performance of the single-mode protocol. All considerations concerning possible improvements of the enhancement factor apply equally to the single-mode DLCZ protocol.
 
\section{Triangular enhancement cavity -- characterization and parameter choice}
 
The cavity was designed taking into account the experimental setup and the properties of the states we wanted to generate. Some of these design considerations are explained in this section.
 
The first aspect is the cavity geometry. In the case of a linear cavity, write photons emitted into each of the two opposite directions defined by the optical mode of the cavity occupy the same spatial mode outside the cavity and will be detected. These two emission directions are linked to spin waves generated  with different wavevector $\bold{k}_{sw}=\bold{k}_W-\bold{k}_w$. Half of the detected write photons correspond to disadvantageous spin waves with a larger wave vector $\bold{k}_{sw}$ which experience a faster decoherence induced by atomic motion \cite{Simon2007c}. This can be solved by using a ring cavity, where emission into two opposite directions couples to two different spatial modes outside the cavity. In particular, we chose a triangular ring cavity geometry. 
 
\begin{figure}[ht]
 	\includegraphics[width=.48\textwidth]{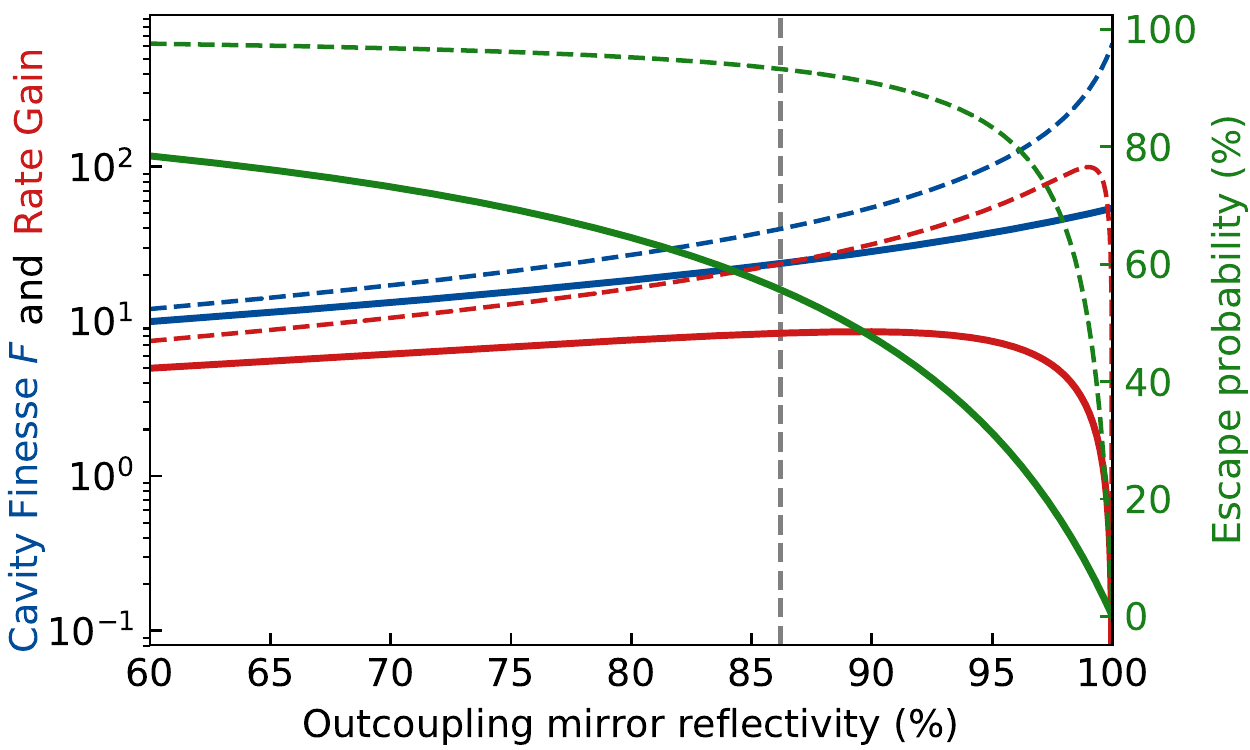}
 	\caption{Cavity finesse, rate gain and escape efficiency as a function of the reflectivity of the cavity outcoupling mirror, assuming total intra-cavity roundtrip losses of 11\% (solid lines, current configuration) and 1\% (dashed lines, possible configuration inside the vacuum chamber). The grey dashed line represents the reflectivity chosen for this cavity.}
 	\label{fig6}
 \end{figure}

An important design parameter is the outcoupling cavity mirror reflectivity $R = 1-T$. Having a high reflectivity increases the finesse of the cavity $F = \pi((1-T)(1-L))^{1/4}/(1-((1-T)(1-L))^{1/2})$, where $L$ are optical roundtrip losses present inside the cavity. This leads to a higher cavity enhancement of the writing. However, increasing the mirror reflectivity also decreases the photon escape efficiency  $\eta_{cav} = T/(T+L)$ of the cavity. The scaling of $F$ and $\eta_{cav}$ is quantified in Fig.~\ref{fig6}. In red, we also plot the expected rate gain for multi-mode cavity operation \cite{Simon2010}, assuming $N_{m} \gg 1$. Allowing for the same multi-mode error in both operating conditions (with and without enhancement), the cavity enhanced memory would offer an up to 8.6-fold increase in coincidence count rate compared to multi-mode operation without cavity.
 
As a compromise, we chose a reflectivity of 86\% (grey dashed line, for H polarisation), corresponding to a measured cavity finesse of 22.2 and a calculated photon escape efficiency of 56\%. Here, we assume a total intra-cavity roundtrip loss of 11\%. Considering that the two MOT chamber windows already account for an $\approx 70 \%$ of these losses, implementing the cavity inside the vacuum chamber will greatly improve the memory performance. Assuming a reasonable roundtrip loss of 1\% for a possible configuration inside the vacuum chamber and better optical elements and polarisation control (dashed lines in Fig.~\ref{fig6}), an up to 99-fold increase in coincidence count rate is within reach.

Another design parameter is the cavity length. Having a shorter cavity leads to a larger free spectral range (FSR) and hence a broader cavity linewidth. Having a broader cavity linewidth allows for the generation of write photons with broader spectrum and hence shorter temporal duration. This means that the repetition rate of the experiment can be higher and more temporal modes can be fit within the memory lifetime. Therefore, the shorter the cavity, the better.  However, due to technical considerations such as geometry of the vacuum chamber setup and spatial filtering of write and read photons with respect to write and read pulses, we choose a cavity length of $\approx$ 85cm. The measured FSR is 342MHz, corresponding to a cavity transmission linewidth of 15.4MHz (inserting the cavity inside the vacuum chamber would also allow for much shorter cavities).

\section{On-demand read-out for entanglement swapping operations}
For application in a quantum repeater node, it is important to retrieve the stored spin wave on demand, in order to provide synchronisation with photons from neighbouring nodes. In our implementation, the retrieval does not happen immediately after requesting the photon. For example, if applied in a DLCZ type quantum repeater link, the gradient can only be reversed upon arrival of a herald from the intermediate measurement station, signalling successful entanglement generation at time $\tau$. Upon gradient reversal at time $\tau + \delta t$, the read photon is then retrieved at time $T_{ret} = 2(\tau + \delta t)$. The read photon can be synchronized with a second read photon from another memory for entanglement swapping by choosing $\delta t$  accordingly. While this requires the storage time of the memory to be twice as large as compared to immediate retrieval, the requirements on storage time are simultaneously relaxed for a multiplexed repeater.

To achieve on-demand read-out without delay, we propose to null the magnetic field gradient after the last mode was stored in the memory. Doing so, the dephasing will be stopped and the atomic phase frozen. Upon an heralding signal, the gradient can be reversed and the spin-wave modes will rephase immediately.This technique could also be implemented using light shifts.

\section{Level scheme of the experiment}

\begin{figure}[ht]
	\includegraphics[width=.48\textwidth]{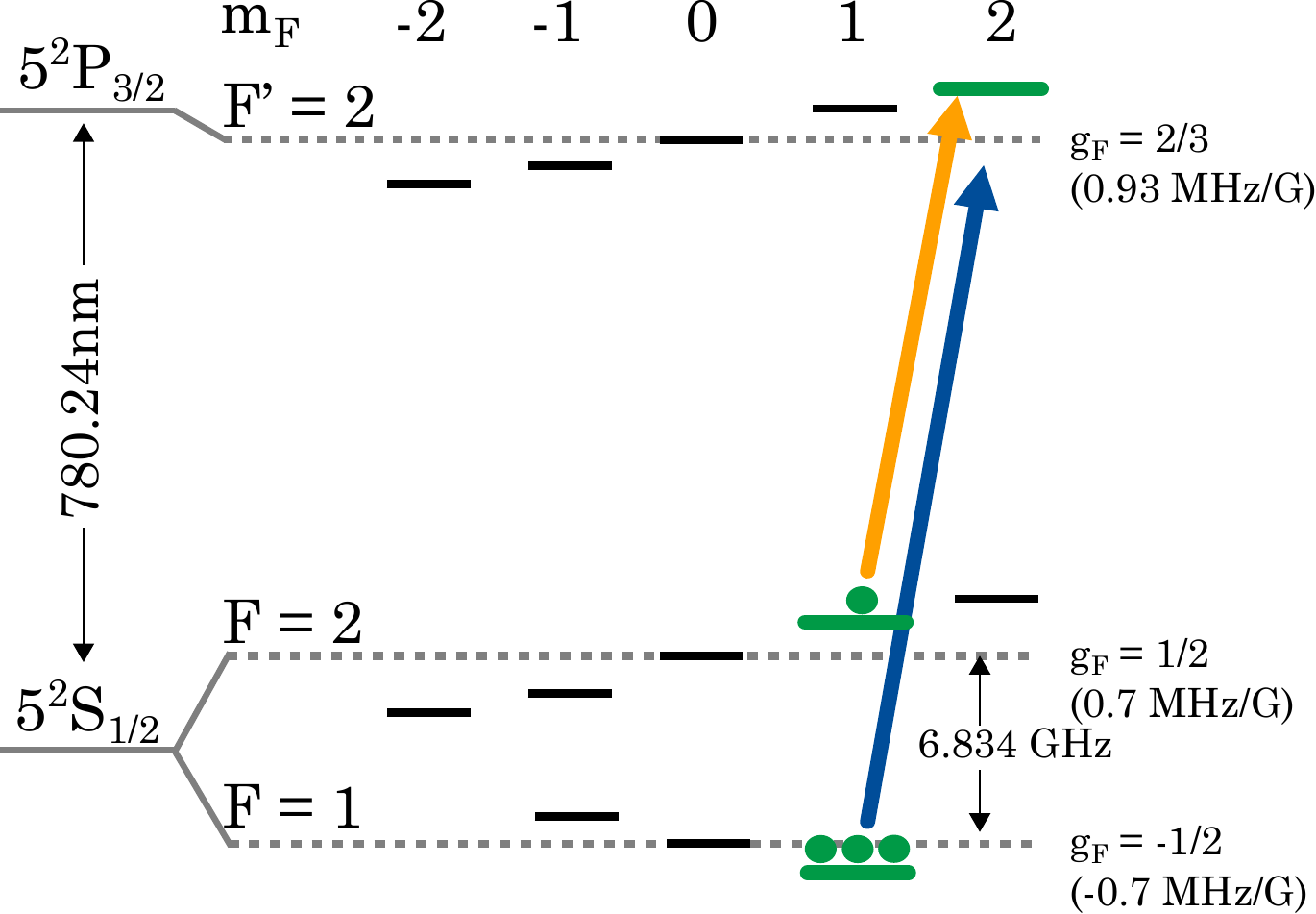}
	\caption{Optical write/read transitions and relevant energy levels on the D2 line of $^{87}$Rb. Blue (yellow) solid lines: write(read) pulses. Blue (yellow) curved lines: read (write) photons.}
	\label{fig7}
\end{figure}

As shown in Fig.~\ref{fig7}, we employ hyperfine transitions on the D2 line of $^{87}$Rb. More specifically, by optical pumping a single Zeeman sublevel $|g\rangle = |5^2S_{1/2},F=1,m_F=1\rangle$ is populated, in the presence of a bias magnetic field aligned with the photon mode. Subsequently, the atoms are off-resonantly exited with $\sigma^+$-polarised light. As we detect photons that are emitted along the field gradient axis only, just a single ($\sigma^+$) write photon transition is allowed for decay to $|5^2S_{1/2},F=2\rangle$, heralding spin wave creation on $|s\rangle = |5^2S_{1/2},F=2,m_F=1\rangle$. As indicated in Fig.~\ref{fig7}, this transition is magnetic-field sensitive, with $\Delta \omega$/B $= 0.7 - (-0.7)$ MHz/G =  1.4 MHz/G. In order to achieve controlled dephasing and rephasing of the spin wave by means of a magnetic gradient field, this dependence is crucial (see Eq. 1 in the main paper). A magnetic field insensitive transition cannot be used for the multi-mode storage protocol.

\section{Deviation from linear increase of coincidence rate with number of modes}

In Fig. 4(a) of the main paper, the gain in coincidence rate $p_{w,r}$ vs. number of modes was investigated. In contrast to the linear increase in spinwave - write photon generation probability $p_w$, $p_{w,r}$ deviated from this linear increase for higher mode numbers. We identify two main effects for this degradation: magnetic field fluctuations and (temporal) uncertainty in feed forward read out.

Spin wave coherence is limited by atomic motion and spurious magnetic gradient fields. As shown in Fig.~\ref{fig5} for the single-mode protocol, this coherence decay decreases the read-out efficiency when the storage time increases. The characteristic 1/e memory lifetime for this operation mode is $\tau \approx 72 \mu s$. However, the multi-mode operation is subject to unstable field conditions due to the presence of additional gradient fields (and field reversal) for the controlled rephasing protocol. This affects the quality of optical pumping to $|g\rangle$ and leads to faster efficiency decay. Moreover, the rephasing time $T_{reph}$ is determined by fulfilling 

\begin{equation}
\int_{0}^{T_{reph}}\Delta w_j(B(t'))dt' = 0.
\label{eqn:6}
\end{equation}

Drifts of the magnetic field strength along a single experimental cycle will lead to rephasing times that are not constant across this cycle. To counteract, we use a high voltage capacitor enabling a fast and stable field reversal. However, this does not allow for removing this gradient drift influence completely for arbitrary storage times, in our current configuration.

While these effect also play a role in Fig. 3(a), the feed forward readout Fig. 4(a) is further affected by the difficulty of reading out a rephased spin wave at its precise rephasing time. Only when reading out at the maximum of the rephasing peak of each mode, one can profit from maximum retrieval efficiency. However, this read out time is programmed beforehand for each mode based on calibration measurements. While the first modes show little jitter in its rephasing time, later modes are subject to greater fluctuations, for example by a change in drift of the above mentioned gradient strength. Reduction of Eddy currents or active field compensation could help mitigating this effect.

In conclusion, additional to the efficiency decay in Fig.~\ref{fig5} for single-mode operation, in multi-mode operation field gradient variations and shifting rephasing times reduce the retrieval efficiency of higher modes above average.